\documentclass[prl,aps,floatfix,twocolumn,tightenlines,nofootinbib,superscriptaddress]{revtex4-1}

\usepackage[utf8]{inputenc}
\usepackage[T1]{fontenc}
\usepackage{graphicx}
\usepackage{amsmath,amssymb,amsthm,amsfonts}
\usepackage{psfrag}
\usepackage{ifsym}
\usepackage{dsfont}
\usepackage{enumerate}
\usepackage{multirow}
\usepackage{hhline}
\usepackage{bbm}
\usepackage[breaklinks=true,colorlinks=true,urlcolor=blue,linkcolor=blue,citecolor=red]{hyperref}
\usepackage{algorithm}
\usepackage{algpseudocode}
\usepackage[dvipsnames]{xcolor}
\usepackage{soul}
\usepackage{siunitx}

\newcommand{\bra}[1] {\langle #1 |}
\newcommand{\ket}[1] {| #1 \rangle}

\newcommand{\ketbra}[1]{ | #1 \rangle\!\langle #1 |}

\newcommand{\sx} {\sigma_\textsc{x}}
\newcommand{\sy} {\sigma_\textsc{y}}
\newcommand{\sz} {\sigma_\textsc{z}}
\newcommand{\su}[1]{\mathrm{SU}(#1)}

\newcommand{\one}{\leavevmode\hbox{\small1\normalsize\kern-.33em1}}

\newcommand{\id}{\mathds{1}}
\newcommand{\Ident} {\mathds 1}
\newcommand{\Tr} {\operatorname{Tr}}

\newcommand{\cinc}{\operatorname{C\textsc{inc}}}
\newcommand{\cnot}{\operatorname{C\textsc{not}}}
\newcommand{\cex}{\operatorname{C\textsc{ex}}}
\newcommand{\ms}{{\operatorname{MS}}}
\newcommand{\R}{{\operatorname{R}}}

\newcommand{\Bra}[1]{{ \langle \! \langle{#1}\vert }}
\newcommand{\Ket}[1]{{ \vert {#1}  \rangle \!  \rangle}}

\newcommand{\BraKet}[2]{{\langle  \!  \langle {#1}\vert {#2} \rangle \! \rangle}}

\newcommand{\Calc}{$^{40}$Ca$^+$ }

\begin{document}
\title{A universal qudit quantum processor with trapped ions}
\author{Martin Ringbauer}
\affiliation{Institut f\"{u}r Experimentalphysik, Universit\"{a}t Innsbruck, Technikerstrasse 25, 6020 Innsbruck, Austria}
\author{Michael Meth}
\affiliation{Institut f\"{u}r Experimentalphysik, Universit\"{a}t Innsbruck, Technikerstrasse 25, 6020 Innsbruck, Austria}
\author{Lukas Postler}
\affiliation{Institut f\"{u}r Experimentalphysik, Universit\"{a}t Innsbruck, Technikerstrasse 25, 6020 Innsbruck, Austria}
\author{Roman Stricker}
\affiliation{Institut f\"{u}r Experimentalphysik, Universit\"{a}t Innsbruck, Technikerstrasse 25, 6020 Innsbruck, Austria}
\author{Rainer Blatt}
\affiliation{Institut f\"{u}r Experimentalphysik, Universit\"{a}t Innsbruck, Technikerstrasse 25, 6020 Innsbruck, Austria}
\affiliation{Institut f\"{u}r Quantenoptik und Quanteninformation, \"{O}sterreichische Akademie der  Wissenschaften, Otto-Hittmair-Platz 1, 6020 Innsbruck, Austria}
\affiliation{Alpine Quantum Technologies GmbH, 6020 Innsbruck, Austria}
\author{Philipp Schindler}
\affiliation{Institut f\"{u}r Experimentalphysik, Universit\"{a}t Innsbruck, Technikerstrasse 25, 6020 Innsbruck, Austria}
\author{Thomas Monz}
\affiliation{Institut f\"{u}r Experimentalphysik, Universit\"{a}t Innsbruck, Technikerstrasse 25, 6020 Innsbruck, Austria}
\affiliation{Alpine Quantum Technologies GmbH, 6020 Innsbruck, Austria}

\begin{abstract}
Today's quantum computers operate with a binary encoding that is the quantum analog of classical bits. Yet, the underlying quantum hardware consists of information carriers that are not necessarily binary, but typically exhibit a rich multilevel structure, which is artificially restricted to two dimensions. A wide range of applications from quantum chemistry to quantum simulation, on the other hand, would benefit from access to higher-dimensional Hilbert spaces, which conventional quantum computers can only emulate. Here we demonstrate a universal qudit quantum processor using trapped ions with a local Hilbert space dimension of up to 7. With a performance similar to qubit quantum processors, this approach enables native simulation of high-dimensional quantum systems, as well as more efficient implementation of qubit-based algorithms.
\end{abstract}

\maketitle

Quantum information processing (QIP) successfully builds on the paradigm of binary information processing that has fueled classical computers for decades. Quantum bits, just like their classical counterparts, are two-level systems with basis states labelled $\ket{0}$ and $\ket{1}$. The underlying physical systems, however, almost always consist of higher-dimensional Hilbert spaces that need to be artificially restricted to fit our binary paradigm. This multi-level structure presents a powerful resource for QIP. Using auxiliary levels as a cache memory for quantum information has enabled the implementation of highly complex quantum algorithms~\cite{Martinez2016} and dissipative processes~\cite{Schindler2013}, as well as improved decomposition of multi-qubit gates~\cite{Lanyon2008}. Yet, experimental realizations in various platforms are limited to proof-of-concept demonstrations~\cite{Senko2015,Ahn2000,Godfrin2017,Anderson2015,Morvan2020,Hu2018a}. Featuring a much richer coherence~\cite{Ringbauer2017} and entanglement structure~\cite{Kraft2018} than their binary counterpart, combined with higher noise resilience~\cite{Cozzolino2019} makes \emph{qudits} (quantum decimal digits) a prime candidate for the next generation of quantum devices.

Here we demonstrate a universal qudit trapped-ion quantum processor (TIQP) using of the native multilevel structure of trapped chains of \Calc ions. Building on the native interactions in the trapped-ion platform, we implement a universal gate set for qudit QIP at a performance approaching that of comparable qubit systems. We further introduce a scalable detection scheme that allows for full qudit readout and develop composite pulse techniques for cross-talk compensation.

\begin{figure}[h!]
    \centering
    \includegraphics[width=\columnwidth]{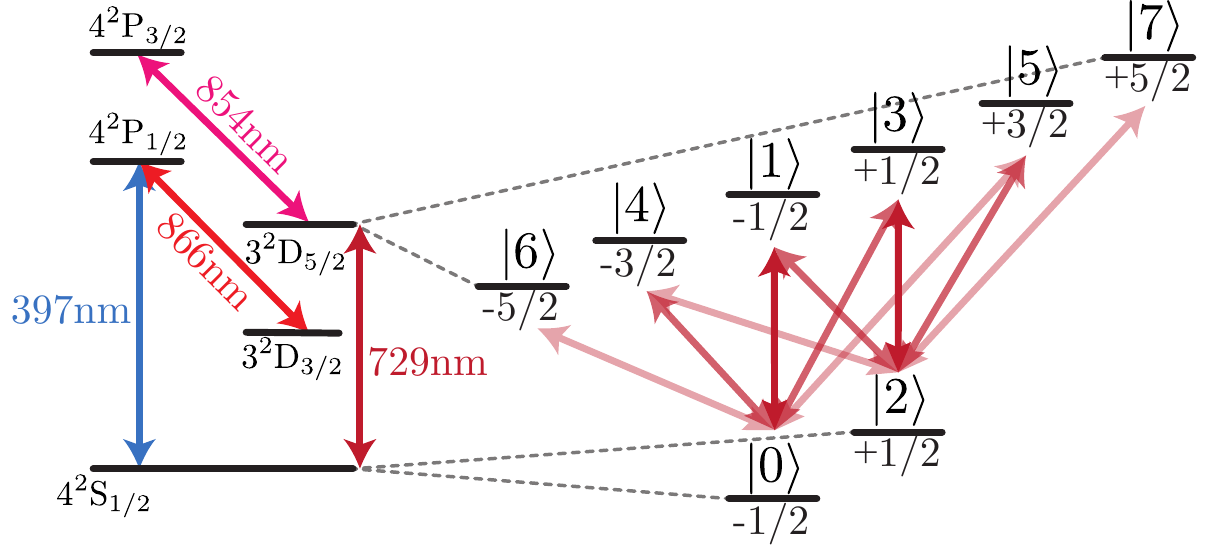}
    \caption{\textbf{Level scheme of the \Calc ion.} Quantum information is encoded in the $S_{1/2}$ and $D_{5/2}$ states, where each transition between S and D is accessible using a single narrowband laser at 729nm.}
    \label{fig.levelscheme}
\end{figure}

Our quantum processor uses a string of \Calc ions in a linear Paul trap~\cite{Schindler2013}. Quantum information is encoded in the $S_{1/2}$ electronic ground state and the metastable $D_{5/2}$ excited state which has a lifetime of $\tau_1 \sim \SI{1.1}{s}$, see Fig.~\ref{fig.levelscheme}. Two sets of permanent magnets generate a magnetic field on the order of $\SI{4}{G}$, which splits the ground state into two Zeeman sublevels ($m=\pm 1/2$) and the excited state into six Zeeman sublevels ($m=\pm 5/2, \pm 3/2, \pm 1/2)$, with a splitting of several MHz. This gives rise to 10 allowed transitions due to the selection rules $\Delta m = 0, \pm 1, \pm 2$, with the coupling strength controlled by the laser polarization and the orientation of the magnetic field. These transitions differ in their sensitivity to magnetic field fluctuations by up to a factor of 5, such that optical qubits are typically encoded in the least sensitive states $\ket{0}=S_{1/2,-1/2}$, $\ket{0}=D_{5/2,-1/2}$. However, with magnetic shielding, spin coherence times on the order of $\SI{100}{ms}$, which is at least 3 orders of magnitude larger than typical gate times, can be achieved for all transitions. Each \Calc ion therefore natively supports a \emph{qudit} with 8 levels, featuring a highly connected Hilbert space, see Fig.~\ref{fig.levelscheme}. 

Ions are initialized close to the electronic and motional ground state by optical pumping and a combination of Doppler cooling, polarization-gradient cooling~\cite{Joshi2020}, and resolved sideband cooling~\cite{Schindler2013}. The native operations available in the TIQP are equatorial rotations on any S-D transition:
\begin{equation}
\R^{i,j}(\theta,\phi)= \exp(-i \theta \sigma_\phi^{i,j}/2) , 
\label{eq:localOps}
\end{equation}
where $\theta$ denotes the rotation angle, $\sigma_\phi^{i,j} = (\cos{(\phi)}{\sx}^{i,j}\pm\sin{(\phi)}{\sy}^{i,j})$ for Pauli matrices $\sx,\sy,\sz$, and $i,j$ indicate the addressed transition according to Fig.~\ref{fig.levelscheme}. Note, that the phase rotation for $\phi$ in $\sigma_\phi^{i,j}$ is positive (negative) when $\ket{i}$ ($\ket{j}$) is in the S-manifold (i.e.\ lower energy). The rotations of Eq.~\eqref{eq:localOps} enable arbitrary $\su{2}$ operations and are thus universal for qubit quantum computing~\cite{Barenco1995} when combined with a two-qubit entangling operation such as the M\o{}lmer-S\o{}rensen gate~\cite{Molmer1999}.
\begin{equation}
\ms^{i,j}(\theta,\phi) = \exp\left(-\frac{i\theta}{4} \left(\sigma_\phi^{i,j}\otimes\Ident + \Ident\otimes \sigma_\phi^{i,j}\right)^2 \right) .
\label{eq:entOps}
\end{equation}
This interaction is achieved by coupling the internal degrees of freedom of the ions to the common motion in the trap using a bichromatic light field. Using multiple focused laser-beams, entangling-gate operations can also be performed on arbitrary subsets of ions in the string, see Supplementary Material (SM) for details. All operations can be performed with error rates below $1\%$ and at least 3 orders of magnitude faster than the coherence time of the system~\cite{Bermudez2017}. Finally, the state of each ion is read out by electron shelving using the short-lived $S_{1/2} \leftrightarrow P_{1/2}$ transition and detecting fluorescence on a CCD camera.

We now demonstrate how to extend the above platform to qudits up to dimension $7$ (one level remains unoccupied for readout). As for qubits, this requires arbitrary single qudit gates, supplemented with one two-qudit entangling gate~\cite{Brennen2006}. For simplicity, we will focus primarily on $d=3$ (qutrits) or $d=5$ (ququints) and outline generalizations to other dimensions.

The set of local operations on a qudit is described by the group $\su{d}$. In the case of a qutrit, the Lie algebra of $\su{3}$ is spanned by the Gell-Mann matrices~\cite{Gell-Mann1962}, which are a natural generalization of the Pauli matrices from $\su{2}$ to $\su{3}$, with further generalizations towards $\su{d}$. The Gell-Mann matrices are a set of traceless, Hermitian matrices, satisfying $\Tr[\lambda_i\lambda_j]=2\delta_{ij}$, where $\delta_{ij}$ is the Kronecker delta:

\begin{align}
    \label{eq:QutritGellMann}
    \lambda_1 &= \begin{pmatrix} 
        0 & 1 & 0 \\
        1 & 0 & 0 \\
        0 & 0 & 0 
    \end{pmatrix} , 
    \lambda_2 = \begin{pmatrix}
        0 & -i & 0 \\
        i & 0 & 0 \\
        0 & 0 & 0 
    \end{pmatrix} , 
    \lambda_3 = \begin{pmatrix}
        1 & 0 & 0 \\
        0 & -1 & 0 \\
        0 & 0 & 0 
    \end{pmatrix} , \nonumber\\
    \lambda_4 &= \begin{pmatrix}
        0 & 0 & 1 \\
        0 & 0 & 0 \\
        1 & 0 & 0 
    \end{pmatrix} , 
    \lambda_5 = \begin{pmatrix}
        0 & 0 & -i \\
        0 & 0 & 0 \\
        i & 0 & 0 
    \end{pmatrix} , \\
    \lambda_6 &= \begin{pmatrix}
        0 & 0 & 0 \\
        0 & 0 & 1 \\
        0 & 1 & 0 
    \end{pmatrix} , 
    \lambda_7 = \begin{pmatrix}
        0 & 0 & 0 \\
        0 & 0 & -i \\
        0 & i & 0 
    \end{pmatrix} , 
    \lambda_8 = \frac{1}{\sqrt{3}}\begin{pmatrix}
        1 & 0 & 0 \\
        0 & 1 & 0 \\
        0 & 0 & -2 \nonumber
    \end{pmatrix} .
\end{align}
Notably, one can identify three independent $\su{2}$ sub-algebras: $\{\lambda_1,\lambda_2,\lambda_3\}$, $\{\lambda_4,\lambda_5,(\lambda_3+\sqrt{3}\lambda_8)/2\}$, $\{\lambda_1,\lambda_2,(-\lambda_3+\sqrt{3}\lambda_8)/2\}$. This reflects the physical implementation in the trapped-ion platform via two-level couplings and makes the Gell-Mann operations a natural starting point for a qudit quantum processor.

Beyond basis operations, an important class of single-qudit gates is the Clifford group, as used in quantum error correction~\cite{Gottesman1998}. Following Ref.~\cite{Clark2006}, we introduce generalized Pauli Z and X operations:
\begin{align}
    Z_d &= \omega_d^j\ket{j}\bra{j} \label{eq:QutritPauliZ} \\
    X_d &= \ket{j+1\text{ (mod }d)}\bra{j} \label{eq:QutritPauliX} .
\end{align}
with $\omega_d = e^{\frac{2\pi i}{d}}$ and $\{\ket{j}\}_{j=0}^{d-1}$ denoting the computational basis in dimension $d$. These operations generate the $d$-dimensional analogue of the Pauli group, from which one constructs the $d$-dimensional Clifford group, with generators
\begin{align}
    H_d &= \frac{1}{\sqrt{d}}\sum_{j,k}\omega_d^{j k} \ket{k}\bra{j} \\
    S_d &= \sum_{j}\omega_d^{j(j+1)/2} \ket{j}\bra{j} .
    \label{eq:QutritClifford}
\end{align}
Achieving universal single qudit operations requires not only Clifford gates, but also one additional non-Clifford gate. A popular choice for such a gate is the ``$\pi/8$'' or ``T'' gate~\cite{Gottesman1998}, which for qutrits is given by~\cite{Howard2012}
\begin{equation}
    T_3= 
    \begin{pmatrix}
        1 & 0 & 0\\
        0 & e^{\frac{2\pi i}{9}} & 0\\
        0 & 0 & e^{-\frac{2\pi i}{9}} .
    \end{pmatrix}
    \label{eq:QutritT}
\end{equation}
In our TIQP any single-qudit operation can be constructed from at most $\mathcal{O}(d^2)$ two-level rotations of Eq.~\eqref{eq:localOps}, via a decomposition into \emph{Givens rotations}~\cite{Brennen2006}, see SM for details. A notable challenge in qudit as compared to qubit QIP is that the elementary two-level operations lose their ``global phase'' gauge freedom. For example, a qubit Z rotation by an angle $\theta$ can be understood as a phase shift $\theta$ applied to $\ket{1}$, or equivalently as phase shifts $\pm\theta/2$ applied to $\ket{0}$ and $\ket{1}$, respectively. In the qudit case, this equivalence breaks down, because any phase shift is measured relative to the spectator level. In the SM we discuss how to overcome this technical challenge when decomposing qudit gates.

\begin{figure}[h!]
    \centering
    \includegraphics[width=0.95\columnwidth]{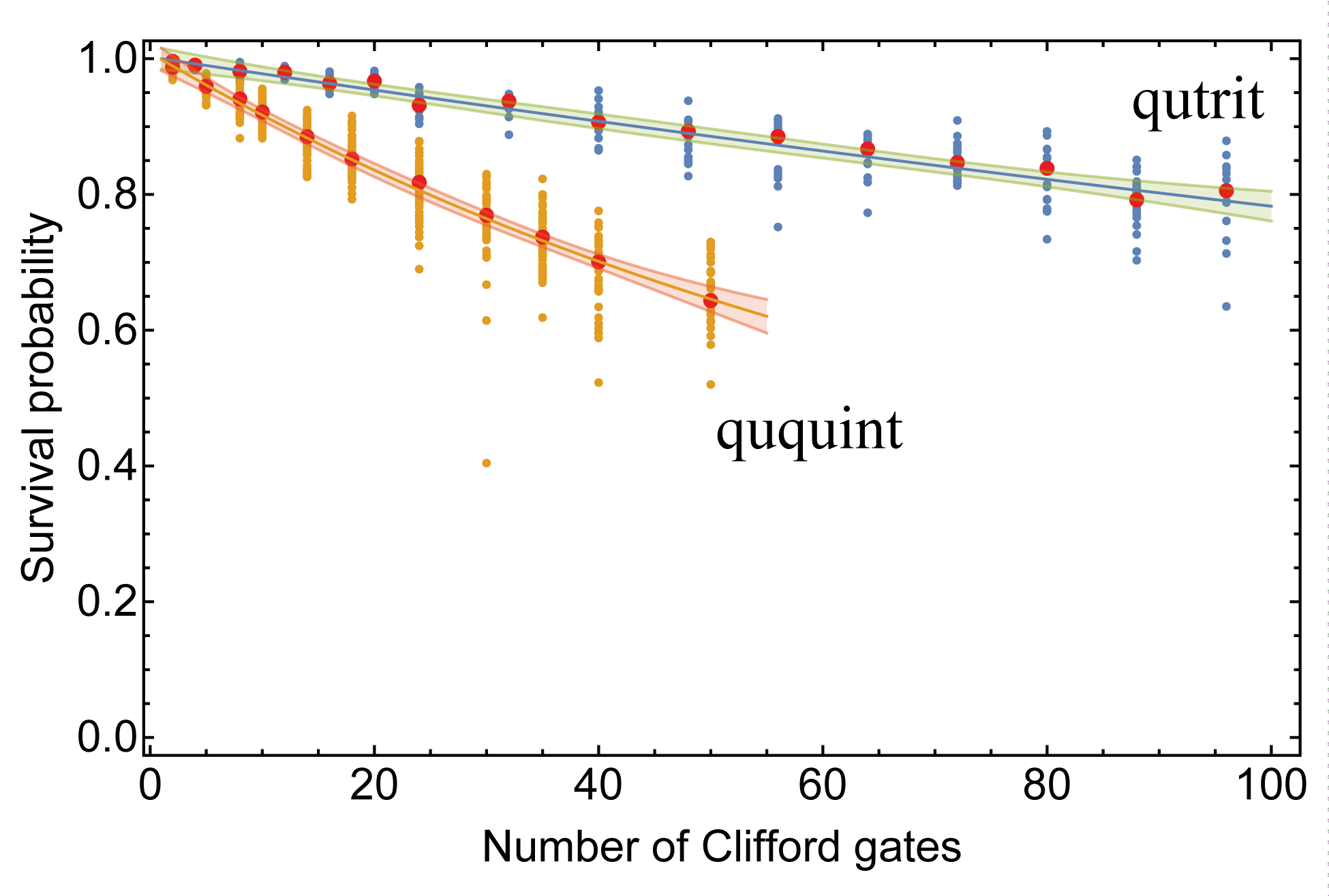}
    \caption{\textbf{Single qudit randomized benchmarking}. From the decay of the survival probability for increasingly long sequences of randomly chosen Clifford gate decompositions of identity, we estimate the average error per gate~\cite{Magesan2012}. For qutrits (blue data points) we obtain an average Clifford error rate of $2(2)\cdot 10^{-3}$ corresponding to an error rate of $2.0^{+0.8}_{-0.5}\cdot 10^{-4}$ per laser pulse. For ququints (orange data points), we obtain an average Clifford error rate of $1.0(2)\cdot 10^{-2}$, corresponding to an error rate of $3.2^{+0.8}_{-0.7}\cdot 10^{-4}$ per laser pulse. For each length we performed at least 20 random sequences (data points), with the median shown in red. The fits (blue and orange lines) and their $99\%$ confidence intervals (shaded regions) are shown as well.}
    \label{fig:RB}
\end{figure}

Experimentally, we demonstrate universal single-qutrit control in a multi-ion crystal by implementing the fundamental Gell-Mann rotations of Eq.~\eqref{eq:QutritGellMann}, as well as the Clifford+T gate set of Eqs.\eqref{eq:QutritPauliX}-\eqref{eq:QutritT}. Process tomography results for each of these operations are shown in the SM. In order to benchmark the performance of our single qudit operations in a way that is not limited by state-preparation-and-measurement (SPAM) errors, we perform randomized benchmarking of the qutrit and ququint Clifford operations~\cite{Magesan2012}, see Fig.~\ref{fig:RB}. For qutrits and ququints we obtain error rates per Clifford operation of $2(2)\cdot 10^{-3}$ and $1.0(2)\cdot 10^{-2}$, respectively. These numbers compare well against the thresholds for quantum error correction, which for qubits are as high as $\sim 1\%$~\cite{Fowler2012} and are expected to improve further with qudit dimension~\cite{Campbell2014}.

Completing the universal toolbox requires at least one qudit entangling gate. The most basic choice here is a two-level controlled-NOT ($\cnot$) gate embedded in a qudit Hilbert space. We call this a controlled-exchange gate, $\cex$, since it exchanges two states $\ket{t_1}, \ket{t_2}$ of the target qudit if and only if the control qudit is in state $\ket{c}$:
\begin{align}
\cex_{c,t_1,t_2} : \begin{cases}
\ket{c,t_1} \leftrightarrow \ket{c,t_2} \\
\ket{j,k} \rightarrow \ket{j,k}  \quad\text{for $j\neq c, k\neq t_1,t_2$} .
\end{cases}
\label{eq:cex}
\end{align}
A slightly generalized version of the $\cex$ is the controlled-increment ($\cinc$) gate~\cite{Brennen2006}, which increments the state of the target qudit by 1, if and only if the control qudit is in state $d-1$:
\begin{equation}
\cinc : \begin{cases}
\ket{j,k} \rightarrow \ket{j,k}  & \text{if $j<d-1$} \\
\ket{j,k} \rightarrow \ket{j,k\oplus 1}  & \text{if $j=d-1$}
\end{cases} ,
\label{eq:cinc}
\end{equation}
where $\oplus$ denotes addition modulo $d$. Other entangling gates, such as the controlled-sum gate ($\mathrm{C}_{\textsc{sum}}\ket{i}\ket{j}=\ket{i}\ket{j+i}$) can be obtained from multiple applications of $\cinc$ or $\cex$.

\begin{figure}[h]
    \centering
    \includegraphics[width=0.95\columnwidth]{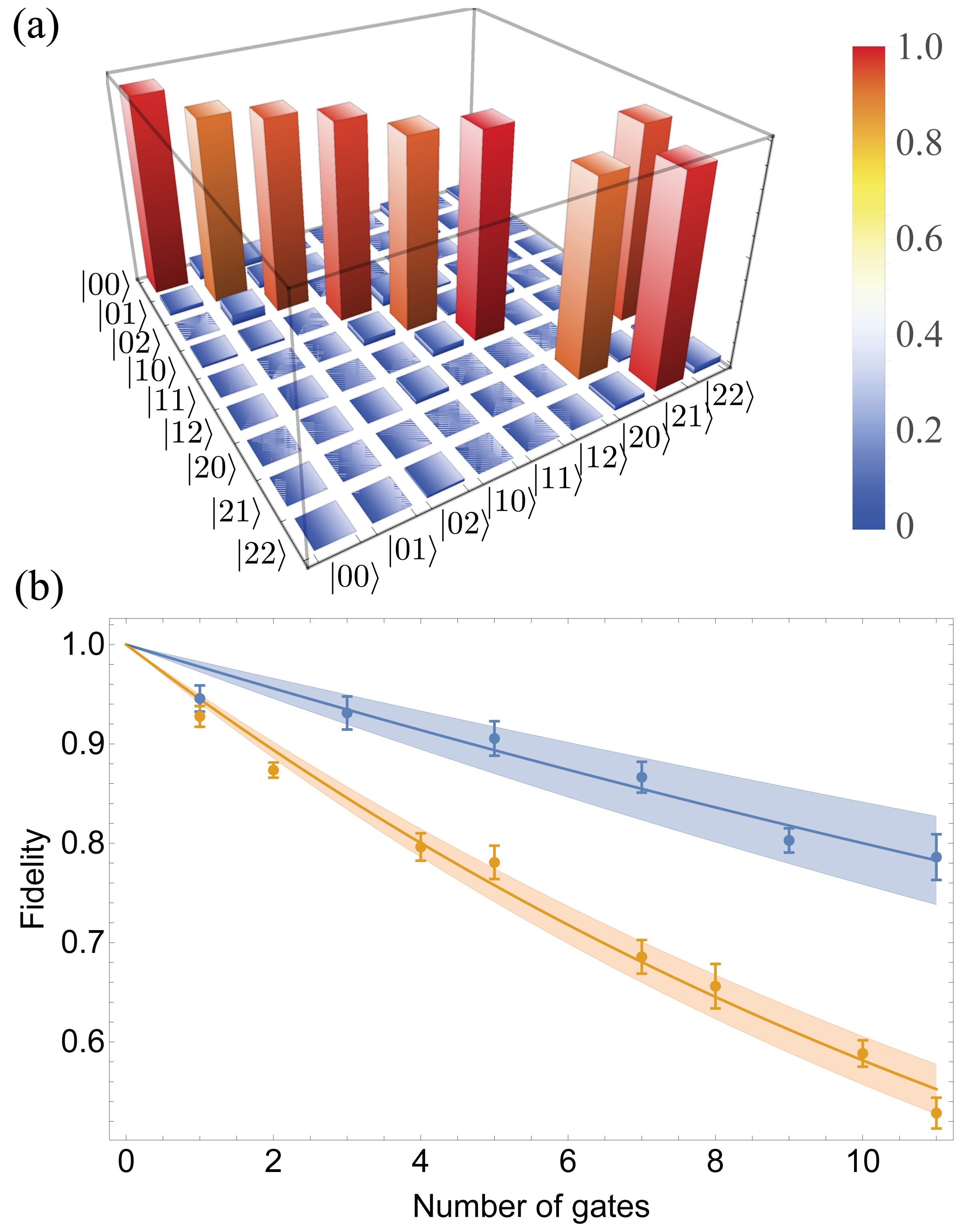}
    \caption{\textbf{Qutrit entangling gates} \textbf{(a)} Experimentally measured truth-table for a controlled-increment gate $\cinc$.
    \textbf{(b)} Fidelity decays for a $\cex$ gate in a qutrit Hilbert space, and a qutrit $\cinc$ gate with initial states $\ket{0+1}\ket{0}/\sqrt{2}$ and $\ket{0+2}\ket{0}/\sqrt{2}$, respectively. The data points correspond to the average of population and coherence at each gate length and the lines are exponential fits from which we extract the average fidelity values. Note that the different number of data points is due to the periodicity of 2 for the $\cex$ gate and 3 for the $\cinc$ gate.}
    \label{fig:CincResults}
\end{figure}

The most straight-forward way to implement two-qudit gates, which we follow here, is again to decompose them into two-level rotations per Eq.~\eqref{eq:localOps} and two-level MS gates per Eq.~\eqref{eq:entOps}. A crucial challenge here is again the ``global'' phase factor $\exp(\mathrm{i} \theta/2)$ of the MS gate, which imparts unwanted phases on spectator states, see SM for details and design considerations. In the case of the $\cex$ gate we find a decomposition into 2 two-level MS gates independent of qudit dimension, and for the $\cinc$ gate we require $2d$ MS gates. Other two-qudit gates can be synthesized in a similar fashion. Alternatively, one could directly generalize the MS gate to a native qudit gate by driving multiple transitions simultaneously. This reduces the gate count, however, only at the cost of increased complexity of the classical control system and more challenging experimental calibration~\cite{Low2020}.

Experimentally, we estimate the fidelity of the $\cex$ and $\cinc$ gates in a SPAM-free way by applying a sequence of entangling gates of increasing length to a fixed input state and measuring both the fraction of runs that end in the correct state (population) and the contrast of the resulting interference fringes (coherence), see Fig.~\ref{fig:CincResults}. Fitting the average of these values we obtain fidelity estimates of $\mathcal{F} = 97.5(2)\%$ and $\mathcal{F} = 93.8(2)\%$ for the $\cex$ and $\cinc$ gates, respectively.

\begin{figure}[ht]
    \centering
    \includegraphics[width=\columnwidth]{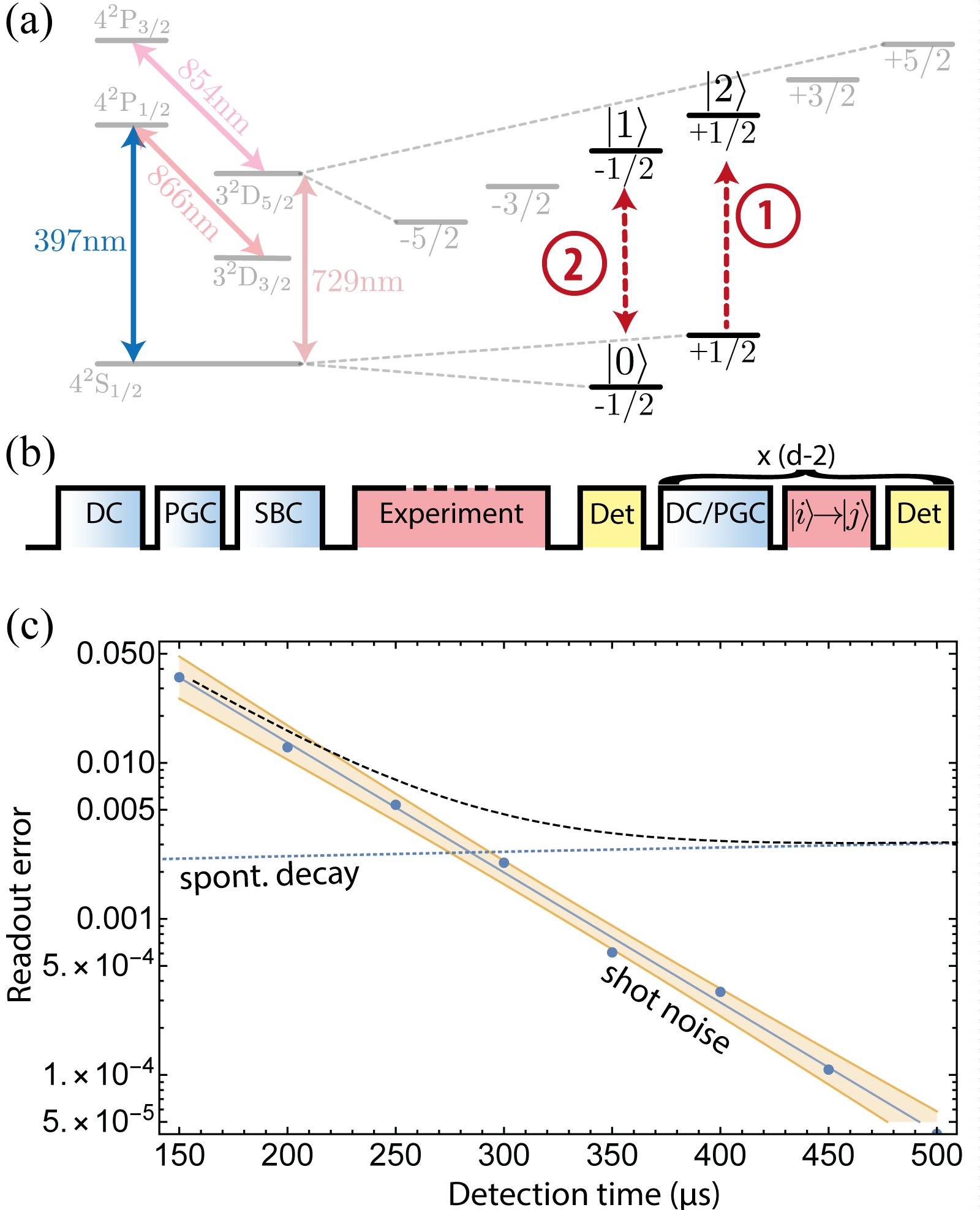}
    \caption{\textbf{(a) Qudit readout} is achieved by successive electron shelving and scattering steps with interleaved recooling. \textbf{(b) Qudit Readout pulse sequence.}
    \textbf{(c) Qudit readout error.} Data points represent measured discrimination uncertainty between bright and dark states at a given measurement time. The solid blue line indicates the fit to this data with $99\%$ confidence regions in orange. The dotted blue line represents the error from spontaneous decay for two detections (qutrit), with $2500\mu s$ of PGC cooling between detections. The combined detection error in the worst case is given by the dashed black line, indicating an optimum of $3\cdot 10^{-3}$ for $\sim 500\mu s$ detection time.}
    \label{fig:readout}
\end{figure}

The final challenge is qudit readout, which requires us to distinguish $d$ different states, whereas conventional fluorescence readout can only distinguish the S and D manifold. To circumvent this problem, we employ a sequential readout scheme. First, we shelve the $\ket{2}$ state to the D manifold (thus one of the 8 levels needs to remain unpopulated). Then we perform a standard fluorescence readout, which projects each ion in the string onto $\{\ketbra{0}, \Ident-\ketbra{0}\}$, allowing us to identify the $\ket{0}$ state without perturbing the rest. This is followed by a reordering pulse, which brings the state $\ket{1}$ into the S manifold and another fluorescence readout. This sequence is repeated until all ions are found in the ``bright'' state at some point in the readout sequence, which lets us assign a unique qudit quantum state to the string of ions. The readout time for this scheme increases only linearly in qudit dimension, see Fig.~\ref{fig:readout}(a).

An additional challenge with longer strings of ions is scattering-induced heating of the ion crystal. This increases the error rate of subsequent reordering pulses and thereby contributes to detection errors. We counter this effect by applying broadband polarization-gradient cooling (PGC) between detection events, see Fig.~\ref{fig:readout}(b). This unavoidably increases the time required for a full readout, leading to a qutrit readout error of $\sim 99.7\%$ for $500\mu s$ detection time and $2500\mu s$ re-cooling time. Note, however, that this is a worst-case error, since those states measured earlier will experience lower spontaneous decay errors than those measured later. Moreover, the readout error can be reduced by an order of magnitude using improved collection optics~\cite{Pogorelov2021} and fast re-cooling techniques~\cite{Schindler2013}.

We demonstrate a universal qudit quantum processor built on state-of-the-art trapped-ion hardware. Our approach requires only a minimal overhead in classical control capabilities and can be integrated into existing experiments. This will particularly benefit applications that are natively formulated in terms of higher spin models, such as quantum chemistry~\cite{MacDonell2020} and quantum simulation of lattice gauge theories~\cite{Rico2018}. Two key challenges for getting the best performance out of qudit hardware are low cross-talk errors due to the large number of local operations required, and fast re-cooling and readout capabilities to reduce SPAM errors. Improvements to the optics and control electronics could improve these numbers by at least an order of magnitude with current technology. Finally, a larger magnetic field would have the benefit of larger spectral separation between the various qudit transition, which will be required for working with longer strings of qudits and the associated crowded mode spectrum. 
Conceptually, a key challenge for making use of the system we present remains in compiling quantum algorithms into the qudit framework. To facilitate such compilation, it will be highly beneficial to complement the embedded two-level entangling gates we presented with a suite of genuine qudit entangling gates.

\begin{acknowledgments}
\noindent\textbf{Acknowledgements}
We thank Marcus Huber and Joel Wallman for discussions. This project has received funding from the European Union’s Horizon 2020 research and innovation programme under the Marie Skłodowska-Curie grant agreement No 840450. It reflects only the author's view, the EU Agency is not responsible for any use that may be made of the information it contains. We also acknowledge support by the Austrian Science Fund (FWF), through the SFB BeyondC (FWF Project No.\ F7109), by the U.S.\ Army Research Office (ARO) through grant no.\ W911NF-21-1-0007, and by the Office of the Director of National Intelligence (ODNI), Intelligence Advanced Research Projects Activity (IARPA), via the U.S. ARO Grant No. W911NF-16-1-0070. This project has received funding from the European Union's Horizon 2020 research and innovation programme under the Marie Sk{\l}odowska-Curie grant agreement No. 801110 and the Austrian Federal Ministry of Education, Science and Research (BMBWF).\\
\noindent\textbf{Author contributions} MR developed the concepts. MR, MM, LP, RS, PS, and TM performed the experiments. MR analyzed the data. TM and RB supervised the project. All authors contributed to writing the manuscript. \\
\noindent\textbf{Competing Interests} The authors declare no competing interests.\\
\noindent Requests for materials and correspondence should be addressed to MR (email: martin.ringbauer@uibk.ac.at).
\end{acknowledgments}

\newpage
\onecolumngrid
\clearpage
\renewcommand{\theequation}{S\arabic{equation}}
\renewcommand{\thefigure}{S\arabic{figure}}
\renewcommand{\thetable}{S\Roman{table}}
\renewcommand{\thesection}{S\Roman{section}}
\setcounter{equation}{0}
\setcounter{figure}{0}
\setcounter{table}{0}
\setcounter{section}{0}
\begin{center}
{\bf \large Supplementary Information: \\
A universal qudit quantum processor}
\end{center}
\medskip

\section{Multi-ion addressing}
\begin{figure}[h!]
    \centering
    \includegraphics[width=\columnwidth]{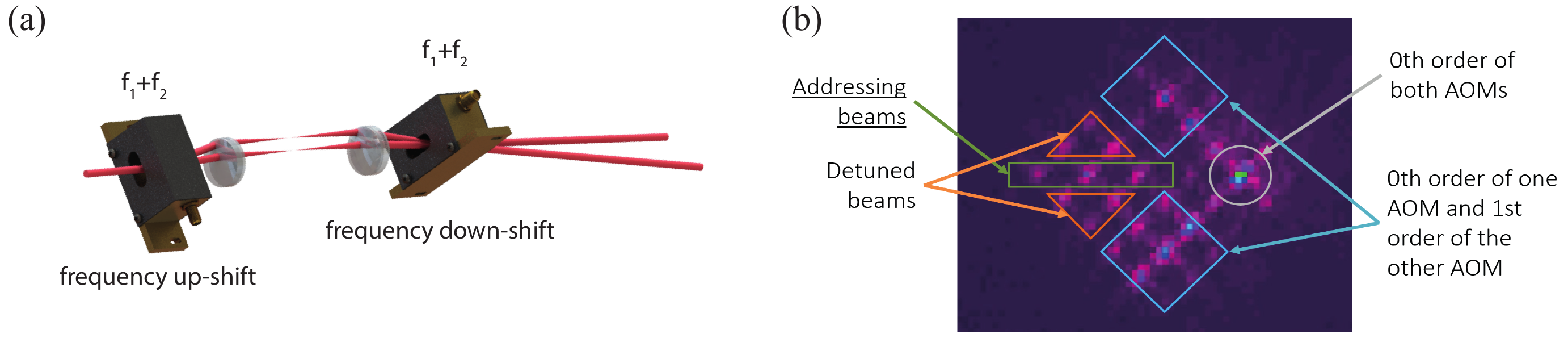}
    \caption{\textbf{Addressing unit for simultaneous multi-ion addressing.} \textbf{(a)} An incoming light beam is frequency up-shifted by first-order diffraction in an AOD rotated $45\deg$ relative to the ion chain. The beam is then imaged into a second AOD rotated $-45\deg$ to the ion chain and frequency down-shifted by the same amount. This results beam that is freely steerable along the ion chain with no frequency shift and full phase control via phase offset between the AODs. Owing to the large bandwidth of the AODs, the addressing range of such a system is only limited by the the subsequent focusing optics into the trap and capable of addressing up to 100 ions.
    \textbf{(b)} Applying $n$ frequencies results in $n$ unshifted spots in the plane of the ions. However, at the same time, there are also $n^2-n$ out-of-plane spots that result from mixing of diffractions corresponding to different frequencies and are thus off-resonant by the difference in addressing frequencies.}
    \label{fig:addressing}
\end{figure}

\begin{figure*}[]
    \centering
    \includegraphics[width=0.9\textwidth]{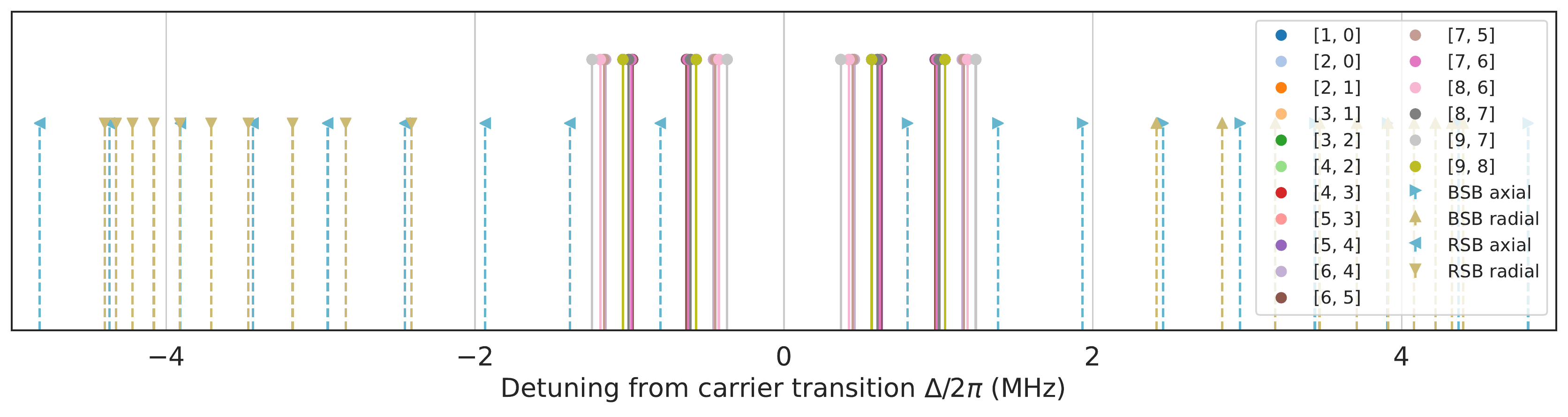}
    \caption{Simulated spectrum of the unwanted off-axis spots (solid stems, integer tuple in the legend denotes the addressed ions) when addressing two ions with a bichromatic light field, slightly detuned from the axial center-of-mass motional sideband, as used for the MS gate. Shown are only the off-axis spots resulting from addressing neighbour and next-neighbour ions, since the others' contributions to cross-talk errors are negligible. Different symbols indicate the frequencies of red (RSB) and blue (BSB) sidebands along the axial and radial direction with respect to the trap axis, see legend. The example is for 10 ions at an axial trap frequency of $\SI{800}{kHz}$, with a displacement ratio of $\sim\SI{65}{kHz/\mu m}$.}
    \label{fig:addressingSpectrum}
\end{figure*}

\begin{figure}[h!]
    \centering
    \includegraphics[width=0.5\columnwidth]{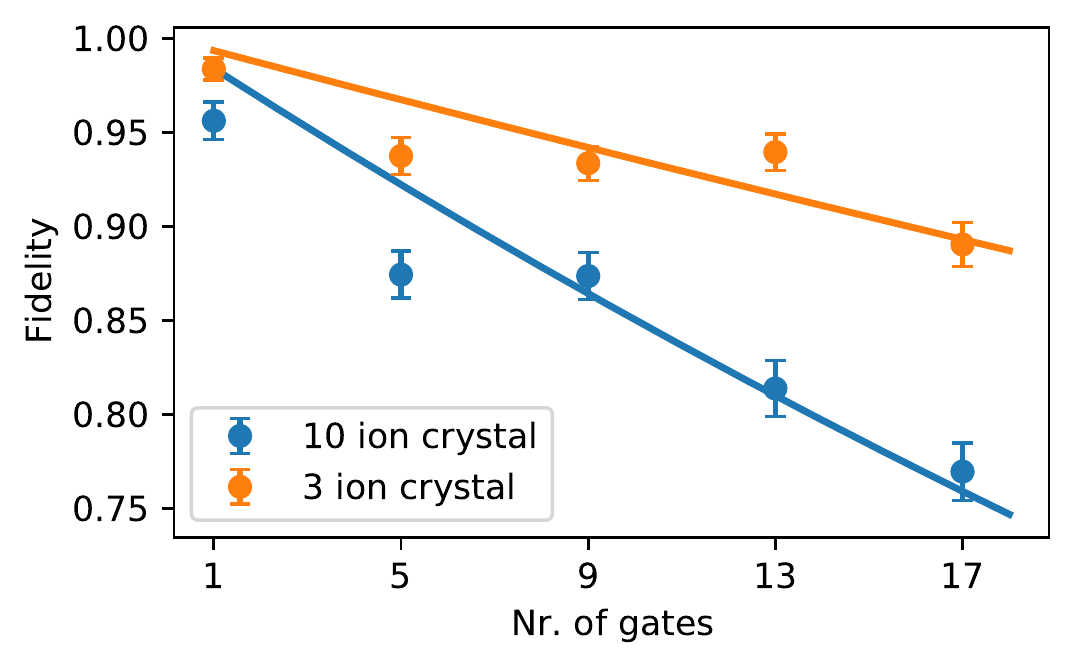}
    \caption{Fidelity decay for addressed entangling gates in a 3 and 10 ion string. The fidelity is $0.993 \pm 0.002$ and $0.983 \pm 0.002$ respectively. Gate performance for ion strings of intermediate length can be interpolated from these values.}
    \label{fig:AddressedMS}
\end{figure}

Individual addressing of ions is achieved with a tightly focused laser beam at an angle of $67.5^\circ$ to the ion string~\cite{Schindler2013}. A combination of two crossed acousto-optic deflectors (AOD) at a $90^\circ$ angle to each other~\cite{kim2008acousto} enables beam steering over a wide range with reconfiguration times below $10\mu s$, see Fig.~\ref{fig:addressing}. By using the positive first order of diffraction on the first AOD and the negative first order on the second AOD, any frequency shifts are exactly cancelled. Multiple ions are addressed by simultaneously applying two (or more) frequencies to the addressing unit. In contrast to approaches with fixed fibre arrays~\cite{Figgatt2019}, our setup allows for precise individual control of each beam to match the non-equidistant ion positions.  Moreover, the phase of each addressed beam can be individually controlled by a phase offset between the RF signals on two AODs for a given driving tone. The price to pay for such flexibility, however, is that $n$ driving tones lead to $n^2$ beams, due to combinations of non-corresponding frequencies of the different AODs, see Fig.~\ref{fig:addressing}. Although beams that originate from mixing of different frequencies are out of the plane of ions, there is residual overlap when addressing ions in close proximity, which leads to cross-talk errors. To avoid such errors, we make use of the fact that any off-axis spots are frequency shifted and thus not resonant with the ion. An important design parameter to ensure that this is true also for bichromatic beams used for entangling gates is the ratio of RF-frequency change to spatial displacement. This parameter can be controlled by the choice of input beam size and focusing optics, which we design for a value of $\sim\SI{65}{kHz/\mu m}$, and a minimal ion separation of $\sim 3.5\mu m$. This ensures that unwanted spots are spectrally detuned from all carriers and motional sidebands for strings up to length 10, see Fig.~\ref{fig:addressingSpectrum}. This optical setup could be further improved by using a beam that is perpendicular to the ion chain and then does not couple to axial motional modes. This system enables two-qubit entangling gates on arbitrary pairs of ions with an error rate that increases only weakly with the number of ions, see Fig.~\ref{fig:AddressedMS}.

\section{Off-resonant operations}
One of the primary error sources in trapped-ion quantum computers is cross-talk from addressed operations~\cite{Parrado-Rodriguez2020}. A common way to combat this problem in conventional TIQPs is through the use of composite pulses, which are a sequence of counter-rotating resonant operations and off-resonant AC-Stark ($\mathrm{Z}$) rotations:
\begin{equation}
    \R(\theta,\phi) = Z(\pi)-\R(-\theta/2,\phi)-Z(\pi)-\R(\theta/2,\phi).
    \label{eq:refocusing}
\end{equation}
The $\mathrm{Z}$ rotations are implemented by a far-detuned laser, which induces energy level shifts due to the AC-Stark effect. Since the Rabi frequency in this case scales quadratically with the resonant Rabi frequency, these pulses benefit from quadratically suppressed cross-talk to neighbouring ions. In the present experiment, the sequence of Eq.~\eqref{eq:refocusing} suppresses cross-talk to $\sim 2\cdot 10^{-3}$ (from $\sim 4\%$ for a resonant operation), which is on the same order of magnitude as the benchmarked gate error rate from Fig.~\ref{fig:RB}. This could be further reduced by several orders of magnitude with improved addressing optics~\cite{Pogorelov2021}.

\begin{figure}[h!]
    \centering
    \includegraphics[width=0.9\columnwidth]{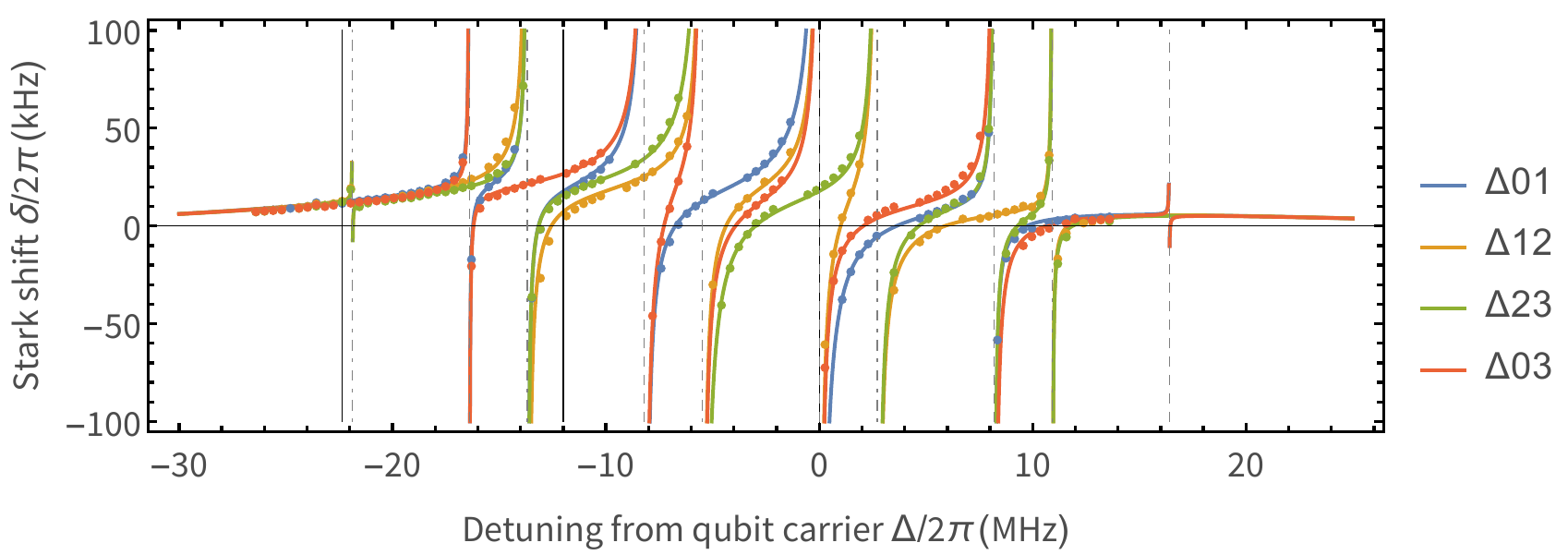}
    \caption{\textbf{AC-Stark shifts in a ququart} Stark shift as a function of the detuning of a single off-resonant beam from the qubit transition for all four transitions in a ququart encoded as in Fig.~\ref{fig.levelscheme}. The solid lines represent the theory prediction according to Eq.~\eqref{eq:Starkshift}, adjusted for the finite bandwidth of the used AOM, where the relative coupling strengths were measured independently. The Rabi frequency, dipole contribution, and AOM bandwidth are fitted to the measured datapoints. Dashed and dot-dashed vertical lines indicate carrier transitions for the $S(-1/2)$ and $S(+1/2)$ states, respectively, and black, solid vertical lines indicate the frequencies of our bichromatic AC-Stark gate to achieve equal Stark shift on the two S and the two D levels. Stark shifts are expected to diverge close to the carrier transitions, and converge to a non-zero background level for large detuning due to coupling to the dipole transitions (not shown).
    }
    \label{fig:StarkShift}
\end{figure}

Reducing cross-talk errors is particularly relevant for qudit quantum processors, as they typically require more basic rotations to realize arbitrary operations than their qubit counterpart. However, when more than two states are occupied, the effect of off-resonant laser interaction is more complex, as each state acquires a different AC-Stark shift due to varying detunings and coupling strengths, see Fig.~\ref{fig:StarkShift}. Formally, the relative shift $\Delta_{i,j}$ between states $\ket{i}$ and $\ket{j}$ is given by~\cite{Haeffner2003}
\begin{equation}
    \Delta_{i,j}(\Delta) = \frac{\Omega^2}{4}\left( 2b - \sum_{t} \frac{\Omega_t^2 \cdot \gamma_{i,j}^{(t)}}{\delta - \delta_t} \right) ,
    \label{eq:Starkshift}
\end{equation}
where the sum runs over all transitions $(t)$, whose coupling strengths relative to the $0\leftrightarrow 1$ transition are given by the dimensionless quantity $\Omega_c$ and can be controlled by the laser polarization. $\delta$ denotes the detuning of the light field, $\delta_t$ is the detuning of transition $t$ from the $\ket{0}\leftrightarrow \ket{1}$ transition, $\Omega$ the Rabi frequency on the $\ket{0}\leftrightarrow \ket{1}$ transition, and $b$ is the Stark shift due to the far-detuned dipole transitions. The relative contributions of the quadrupole transitions to the Stark-shift obtained by the $\ket{i}\leftrightarrow \ket{j}$ transition is given by $\gamma_{i,j}$, which takes the value 2 for the transition $\ket{i}\leftrightarrow\ket{j}$, 1 for transitions that include either $\ket{i}$ or $\ket{j}$, and 0 otherwise.

From Eq.~\eqref{eq:Starkshift} and Fig.~\ref{fig:StarkShift} it is clear that, in the qudit case, a single off-resonant light field would lead to practically uncontrollable phase shifts outside the addressed transition. However, a careful analysis shows that for a system with $d$ occupied levels, a multi-chromatic off-resonant light field with $(d-1)$ tones can, in principle, exactly compensate all level shifts, but one. This provides an avenue for targeted phase shifts of any level in a qudit, enabling composite pulse techniques like in Eq.~\eqref{eq:refocusing}. However, each such beam contains $2\cdot(d-2)$ coupled parameters, which can make their experimental calibration challenging.
\begin{equation}
\mathrm{Z}^i(\theta)\ket{j} = \begin{cases}
e^{-i \theta}\ket{j}  & \text{if $j=i$} \\
\ket{j}  & \text{otherwise}
\end{cases} .
\label{eq:QuditStark}
\end{equation}

\begin{figure}[ht]
    \centering
    \includegraphics[width=0.95\columnwidth]{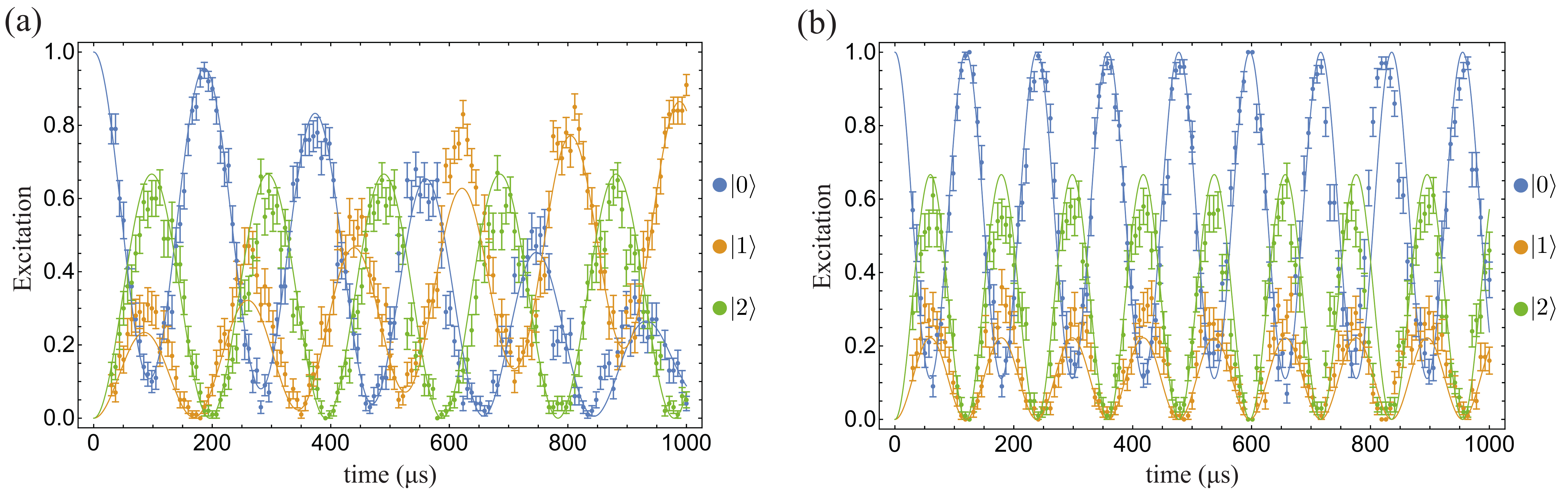}
    \caption{\textbf{Qudit AC-Stark pulses.}
    To verify the performance of our qudit off-resonant operations, we perform a Ramsey experiment, where we create an equal superposition of all qutrit states, followed by evolution under a \textbf{(a)} single-tone, \textbf{(b)} compensated two-tone AC-Stark pulse, followed by a reversal of the first operation. Solid lines correspond to a fit, points are measured data with one standard deviation statistical uncertainties.
    }
    \label{fig:QuditAC}
\end{figure}

Fortunately, a simpler approach suffices for our purpose, where we do not need arbitrary phase shifts on individual levels, but merely want to generalize the composite pulse sequence of Eq.~\eqref{eq:refocusing} to qudits. In this case, it is sufficient to limit the Stark shifts of all the levels to multiples of $\pi$, rather than nulling all but one. For example, it is always sufficient to equally shift the ``S'' levels $\ket{0}, \ket{2}$ in Fig.~\ref{fig.levelscheme} with respect to the ``D'' levels, since shifts on spectator levels cancel by construction. Experimentally, we demonstrate such composite pulses for qutrits and ququarts using a bichromatic off-resonant light field with one frequency far detuned at $\SI{-10}{MHz}$ and a second frequency at $\SI{-0.5}{MHz}$ from the $0\leftrightarrow 6$ transition, see Fig.~\ref{fig:QuditAC}.

\section{Single-qudit gate decomposition}
Recall that the native single-ion operations available in the ion trap are equatorial rotations on any S-D transition, except those with $\Delta m = \pm 3$, given by
\begin{equation}
\R_{i,j}(\theta,\phi) = \exp(-i \theta (\cos{(\phi)}{\sx}^{i,j} \pm \sin{(\phi)}{\sy}^{i,j})/2) .
\label{eq:SUPPlocalOps}
\end{equation}
Here $\sigma_{x,y,z}$ are the Pauli X,Y,Z matrices, $\theta$ denotes the rotation angle, and $i,j$ with $i<j$ indicate the addressed transition. Recall, that the phase convention for $\phi$ in $\sigma_\phi^{i,j}$ is positive when $\ket{i}$ is in the S-manifold and negative when $\ket{j}$ is in the S-manifold.

We now discuss how arbitrary $SU(d)$ operations can be decomposed into the basic rotations of Eq.~\ref{eq:SUPPlocalOps}, adapted from Ref~\cite{OLeary2006a}. Following Ref.~\cite{OLeary2006a} the idea is to first bring the target unitary into diagonal form using the subspace rotations of Eq.~\ref{eq:SUPPlocalOps}, and then implement appropriate phase-shifts between the qudit states. For the latter step it is in general sufficient to implement the target unitary up to a global phase $\gamma$, which requires at most $d-1$ phase gates between pairs of levels, which are decomposed into 3 resonant operations each. Importantly, this procedure works for any coupling map between the qudit levels, although the latter might affect the efficiency of the decomposition. In the case of ladder-type coupling, algorithm~\ref{alg:SUPP_localdiag} provides a decomposition of a generic unitary $U\in \operatorname{SU}(d)$ (indices run from $1$ to $d$). In case of other coupling structures, the same algorithm can be used, albeit with an adapted sequence of operations. Note that the algorithm returns a sequence of gates that must be applied in reverse order and conjugate-transposed for implementing the corresponding unitary in the lab.

\begin{figure}[b]
\begin{algorithm}[H]
\caption{Decomposition of $U\in \operatorname{SU}(d)$ into two-level rotations adapted from Ref.~\cite{OLeary2006a}}
\label{alg:SUPP_localdiag}
    \begin{algorithmic}
    \State $\tilde U = U$
    \State $\operatorname{decomp} = \{\}$
    \For{$c=1, c<d$}
        \For{$r=d, r>c$}
            \If{$\tilde U_{r,c}\neq 0$}
                \State $\theta = 2\arctan[|\tilde U_{r,c}/ \tilde U_{r-1,c}|]$
                \State $\phi = -(\arg[\tilde U_{r-1,c}] - \arg[\tilde U_{r,c}])$
                \State $\tilde U = \operatorname{R}(\theta,\phi)_{r,c} \cdot \tilde U$
                \State $\operatorname{decomp}.\operatorname{append}(\operatorname{R}(\theta,\phi)_{r,c})$
                \State $r = r-1$
            \EndIf
        \EndFor
        \State $c = c+1$
    \EndFor
    \State 
    \State \textbf{Solve} $\begin{pmatrix}
        1 & 0 & 0 & \cdots & 0 \\
        -1 & 1 & 0 & \cdots & 0 \\
        \vdots  & \vdots & \vdots  & \ddots & \vdots \\
        0 & 0 & 0 & \cdots & -1 
        \end{pmatrix} \cdot 
        \begin{pmatrix} \gamma_1/2 \\ \gamma_2/2 \\ \vdots \\ \gamma_{d-1}/2 \end{pmatrix}
        + 
        \begin{pmatrix} \gamma \\ \gamma \\ \vdots \\ \gamma \end{pmatrix}
        = 
        \arg[\operatorname{diag}(\tilde U)]$
    \For{$i=1, i<d$}
        \If{$\gamma_i \neq 0$}
        \State $\operatorname{decomp}.\operatorname{append}(\operatorname{R}(\pi/2,\pi)_{i,i+1} \cdot \operatorname{R}(\gamma_i,\pi/2)_{i,i+1} \cdot \operatorname{R}(\pi/2,0)_{i,i+1})$
        \EndIf
    \EndFor
    \State \Return $\operatorname{decomp}$
   \end{algorithmic}
\end{algorithm}
\end{figure}

\section{Two-qudit gate synthesis}
\begin{figure}[!h]
    \centering
    \includegraphics[width = 0.6\textwidth]{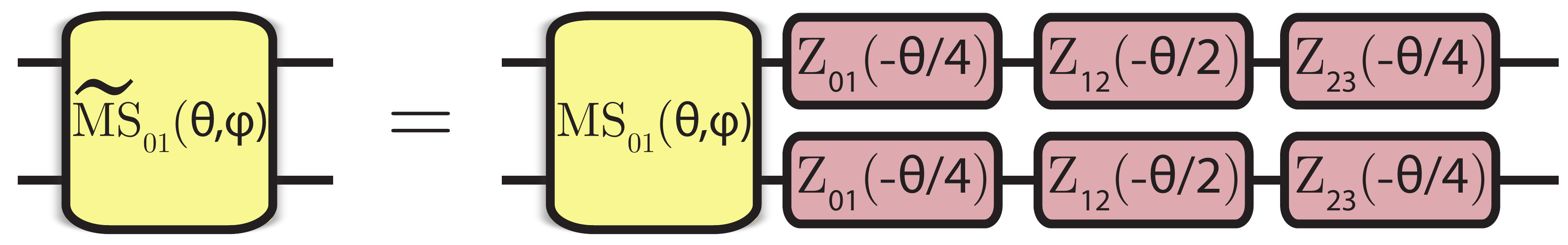}
    \caption{Phase-compensated MS gate on the 01-qubit subspace in a 4-level system. For qudits with other dimensions, the procedure is similar.}
    \label{fig:Supp:phasecorrectedMS}
\end{figure}

We now sketch a strategy for synthesising typical two-qudit gates using the standard gate set outlined in the main text. The two main challenges here are to control off-target operations (i.e.\ operations beyond the addressed subspace) and undesired (``global'' or local) phases of the employed operations. To this end, the first gadget we define is a phase-compensated MS-gate, denoted $\widetilde{\mathrm{MS}}_{i,j}(\theta,\phi)$, see Fig.\ref{fig:Supp:phasecorrectedMS}. Here we employ additional phase-gates between all levels of the qudit to cancel unwanted phases of the MS-gate operation, such that it fully acts as the identity outside (0,1)-subspace that we want to address. These phase-gates will largely be absorbed into other local operations in the final compilation step.

\subsection{Decomposition of the $\cex$ gate}
Using this phase-compensated gate, we now construct the most basic qudit gate, an embedded $\cnot$ gate acting on some two-level subspace of two qudits. We denote this gate a controlled-exchange $\cex$ gate. A key observation at this point is that the (phase-compensated) MS gate acts non-trivially if and only if both qudits are within the correct two-level subspace. As soon as one qudit is outside this subspace, the gate acts as the identity operation. As a consequence, the standard transformation between MS- and $\cnot$-gate cannot be used for our purpose, since it involves local operations that do not correspond to a basis change. Hence in the cases where one of the qudits is outside the addressed subspace and the MS-gate acts as the identity, we are left with undesired local rotations, which effectively induce $\sqrt{\cnot}$-type couplings.

\begin{figure}[!h]
    \centering
    \includegraphics[width = 0.85\textwidth]{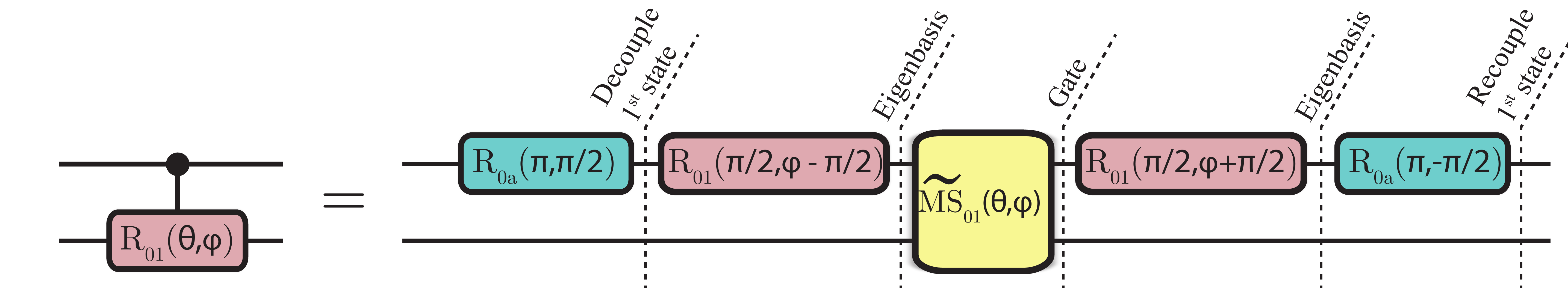}
    \caption{Turning the MS-gate into a controlled rotation by a combination of local basis transformation and spectroscopic decoupling.}
    \label{fig:Supp:MScrot}
\end{figure}

We overcome this problem by exploiting spectroscopic decoupling to turn the MS-gate directly into a controlled-$\operatorname{R}(\theta,\phi)$-rotation. First, rotating the \emph{control} ion into the eigenbasis of the MS-gate, the gate imparts a conditional local operation on the \emph{target} ion as
\begin{align}
(\R_{0,1}(\pi/2,\phi + \pi/2) \otimes \Ident) \cdot
\widetilde{\ms}_{0,1}(\theta,\phi) \cdot 
(\R_{0,1}(\pi/2,\phi - \pi/2) \otimes \Ident) &= \nonumber\\
e^{-\mathrm{i} \theta/2} \left(
\ketbra{0} \otimes \operatorname{R}_{0,1}(-\theta,\phi) + 
\ketbra{1} \otimes \operatorname{R}_{0,1}(\theta,\phi) \right) & ,
\label{eq:Supp:MSdecoupled}
\end{align}
and identity on all other qudit states. Second, we need to drop the unwanted rotation in case the control ion is in state $\ket{0}$, which, as before, cannot be done using local operations. We achieve this, by decoupling the state $\ket{0}$ of the control ion from the interaction by transferring it to an auxiliary qudit level $\ket{a}$, see Fig.\ref{fig:Supp:MScrot}. Using this decomposition, a $\cex$-gate on any two-level subspace of a qudit can be achieved using one $\ms(\pi)$ gate, compared to one $\ms(\pi/2)$ gate required for the standard qubit decomposition.

\subsection{Decomposition of the $\operatorname{C_{INC}}$ gate}
\begin{figure*}[h!]
    \centering
    \includegraphics[width=\columnwidth]{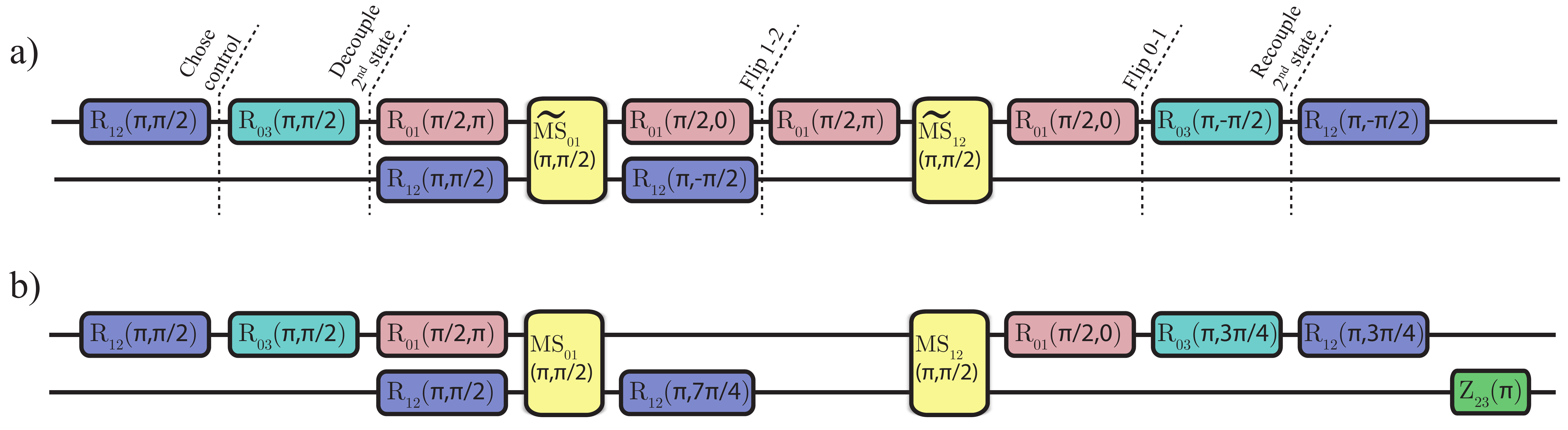}
    \caption{\textbf{Decomposition of a qutrit $\cinc$-gate into two-level operations.} Operations acting on different transitions are color-coded as a visual guide. \textbf{a)} Simple decomposition into controlled rotations using phase-compensated MS-gates. \textbf{b)} Fully compiled decomposition where phase-gates have been absorbed into frame-changes of the local operations. One remaining $\operatorname{Z}$ operation is performed in the standard Euler-decomposition as $\operatorname{Z}_{2,3}(\theta) = \R_{2,3}(\pi/2,-\pi/2) \cdot \R_{2,3}(\theta,0) \cdot \R_{2,3}(\pi/2,\pi/2)$.}
    \label{fig:supp:cinc}
\end{figure*}

As the qudit generalization of the $\cnot$-gate, the $\cinc$-gate is a qudit $\operatorname{X}$ operation, controlled on the state $\ket{d}$ of the control qudit. In the case $d=3$, we have $\operatorname{X}_3 = \R_{0,1}(\pi,\pi/2) \cdot  \R_{1,2}(\pi,\pi/2)$. Using the above strategy, we replace the local rotations in the decomposition of $\operatorname{X}_3$ by the controlled rotations of Fig.~\ref{fig:Supp:MScrot}. For experimental convenience, we use additional local $\pi$-pulses to implement all $\ms$-gate operations in the $(0,1)$-subspace. Finally, we replace the phase-corrected $\widetilde{\ms}$-gates by standard $\ms$-gates and absorb the remaining phase shifts, where possible as frame-rotations of the local gates, see Fig.~\ref{fig:supp:cinc}. This decomposition achieves a $d$-dimensional $\cinc$ gate, using $2(d-1)$ pairwise fully entangling gates, see Fig.~\ref{fig:supp:cinc}.

\newpage

\section{Qudit Tomography}
In order to reconstruct qudit states and processes, we employ a semi-definite programming approach that is independent of the underlying dimension and compatible with any (potentially incomplete) set of experimental data. For the sake of simplicity, we start with state tomography and outline the straightforward generalization to process tomography later.

Following the notation of Ref.~\cite{Ringbauer2015}, let $\mathcal{X}_d \cong \mathbb C^d$ be a $d$-dimensional complex Hilbert space, and $D(\mathcal X_d)$ the set of density matrices on $\mathcal X_d$. We perform measurements using a set of projectors $\{\Pi_i\}_{i=1}^K$, where $\Pi_i=\ketbra{\psi_i}$ to reconstruct an unknown qudit state $\sigma\in D(\mathcal{X}_d)$. The probability for observing a click for projector $\Pi_i$ when the qudit state $\sigma$ was prepared is then given by 
\begin{equation}
p_{i} = \Tr\left[\Pi_i \sigma \right] = \BraKet{\Pi_{i}}{\sigma} .
\label{eq:ProbsQST}
\end{equation}
Here $\Ket{\sigma}=\sum_{i,j=0}^{d-1} \sigma_{i,j}\ket{j}\otimes\ket{i}$ denotes the column vector obtained by stacking the columns of the operator $\sigma$. We now define the vector of observed frequencies $\ket{f}$, the quadratic form $S$, and the weight matrix $W$ as follows:
\begin{equation}
\ket{f} = \sum_{i=1}^K \frac{n_j+\beta}{N_j+d\beta}\ket{i}	\qquad\qquad
S = \sum_{i=1}^K \ket{i}\Bra{\Pi_{i}} \qquad\qquad
W = \sum_{i=1}^K \sqrt{\frac{N_i}{f_i(1-f_i)}} \ketbra{i} ,
\label{eq:TomoDefsQST}
\end{equation}
where $N_j$ is the number of measurement runs in which projector $\Pi_j$ was measured, and $n_j$ is the number of observed counts. The weight matrix $W$ takes into account the multinomial distribution of the observed frequencies, and $\beta$ is an optional hedging parameter, typically set to $\beta\sim 0.5$ to avoid probabilities of 0 or 1 \cite{Ferrie2012Hedged}. Using this, we can now perform a maximum likelihood reconstruction of the state $\sigma$ by solving the following convex optimization problem
\begin{align}
&\text{minimize}\quad \| W(S\Ket{\sigma}-\ket{f}) \|_2
\nonumber\\
&\text{subject to: } \quad\sigma \ge 0 , \Tr[\sigma]= 1 .
\label{eq:MLEopt}
\end{align}
Experimentally, an informationally complete set of the projectors $\{\Pi_i\}_{i=1}^K$ is constructed by combining several orthonormal bases of the Hilbert space. In the case $d=2$ these are the given by the three Pauli operators. In larger dimensions, we similarly construct them out of pair-wise superpositions, as these are most convenient experimentally. For $d=3$ we have, for example, 6 bases, defined by the following collection of pure states (or projectors):
\begin{align}
    &\{ (\ket{0}+i\ket{1})/\sqrt{2}, (\ket{0}-i\ket{1})/\sqrt{2}, \ket{2} \}, \nonumber\\
    &\{ (\ket{0}+\ket{1})/\sqrt{2}, (\ket{0}-i\ket{1})/\sqrt{2}, \ket{2} \}, \nonumber\\
    &\{ (\ket{0}+i\ket{2})/\sqrt{2}, \ket{1}, (\ket{0}-i\ket{2})/\sqrt{2} \}, \nonumber\\
    &\{ (\ket{0}+\ket{2})/\sqrt{2}, \ket{1}, (\ket{0}-\ket{2})/\sqrt{2} \}, \nonumber\\
    &\{ (\ket{1}+i\ket{2})/\sqrt{2}, \ket{0}, (\ket{1}-i\ket{2})/\sqrt{2} \}, \nonumber\\
    &\{ (\ket{1}+\ket{2})/\sqrt{2}, \ket{0}, (\ket{1}-\ket{2})/\sqrt{2} \},
\label{eq:QutritBasis}
\end{align}

In the case of quantum process tomography, we aim to reconstruct the Choi operator $\Lambda_{\mathcal E}$ of a quantum process $\mathcal E$, such that
\begin{equation*}
    \mathcal{E}(\rho) = \Tr_{1}[(\rho_i^T\otimes\id_{\mathcal X_2})\Lambda_{\mathcal{E}}] .
\end{equation*}
This is achieved by preparing an informationally complete set of input states $\{\rho_i\}_{i=1}^K$ and for each of them measuring an informationally complete set of projectors $\{\Pi_j\}_{j=1}^K$. In practice, we chose these two sets to be the same. The probability for observing a click for projector $\Pi_j$ when the state $\rho_i$ was subject to the process $\mathcal{E}$ is then given by
\begin{equation}
p_{ij} = \Tr\big[\Pi_j \Tr_{1}[(\rho_i^T\otimes\id)\Lambda]\big] = \Tr\big[(\rho_i^T\otimes\Pi_j)\Lambda\big]	
	= \BraKet{\Pi_{ij}}{\Lambda} ,
\label{eq:ProbsQPT}
\end{equation}
where $\Pi_{ij} \equiv \rho_i^\ast\otimes\Pi_j$. With this we straightforwardly generalize Eq.~\eqref{eq:TomoDefsQST} as follows:
\begin{equation}
\ket{f} = \sum_{i,j=1}^K \frac{n_{ij}+\beta}{N_{ij}+d\beta}\ket{i,j}	\qquad\qquad
S = \sum_{i,j=1}^K \ket{i,j}\Bra{\Pi_{ij}} \qquad\qquad
W = \sum_{i,j=1}^K w_{ij} \ketbra{i,j} .
\label{eq:TomoDefsQPT}
\end{equation}
Equation~\eqref{eq:MLEopt} then becomes
\begin{align}
&\text{minimize}\quad \| W(S\Ket{\Lambda_{\mathcal{E}}}-\ket{f}) \|_2
\nonumber\\
&\text{subject to: } \quad\Lambda_{\mathcal{E}} \ge 0 , \Tr[\Lambda_{\mathcal{E}}]= d .
\label{eq:MLEoptQPT}
\end{align}

\begin{figure}[h!]
    \centering
    \includegraphics[width=\columnwidth]{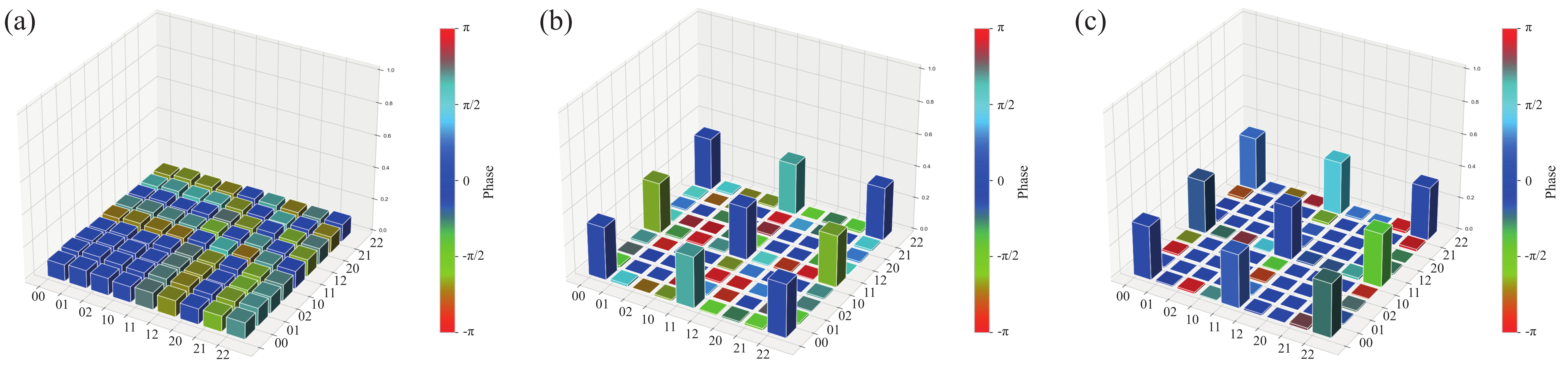}
    \caption{\textbf{Single qutrit Clifford + T gate set}. Shown are experimentally reconstructed Choi operators corresponding to the Clifford gates\textbf{(a)} $H_3$ and \textbf{(b)} $S_3$, as well as the non-Clifford gate \textbf{(c)} $T_3$.The fidelities are given in Tab.~\ref{tab:single_qudit_results}.}
    \label{fig:Supp_LocalClifford}
\end{figure}

\begin{figure}[h!]
    \centering
    \includegraphics[width=\columnwidth]{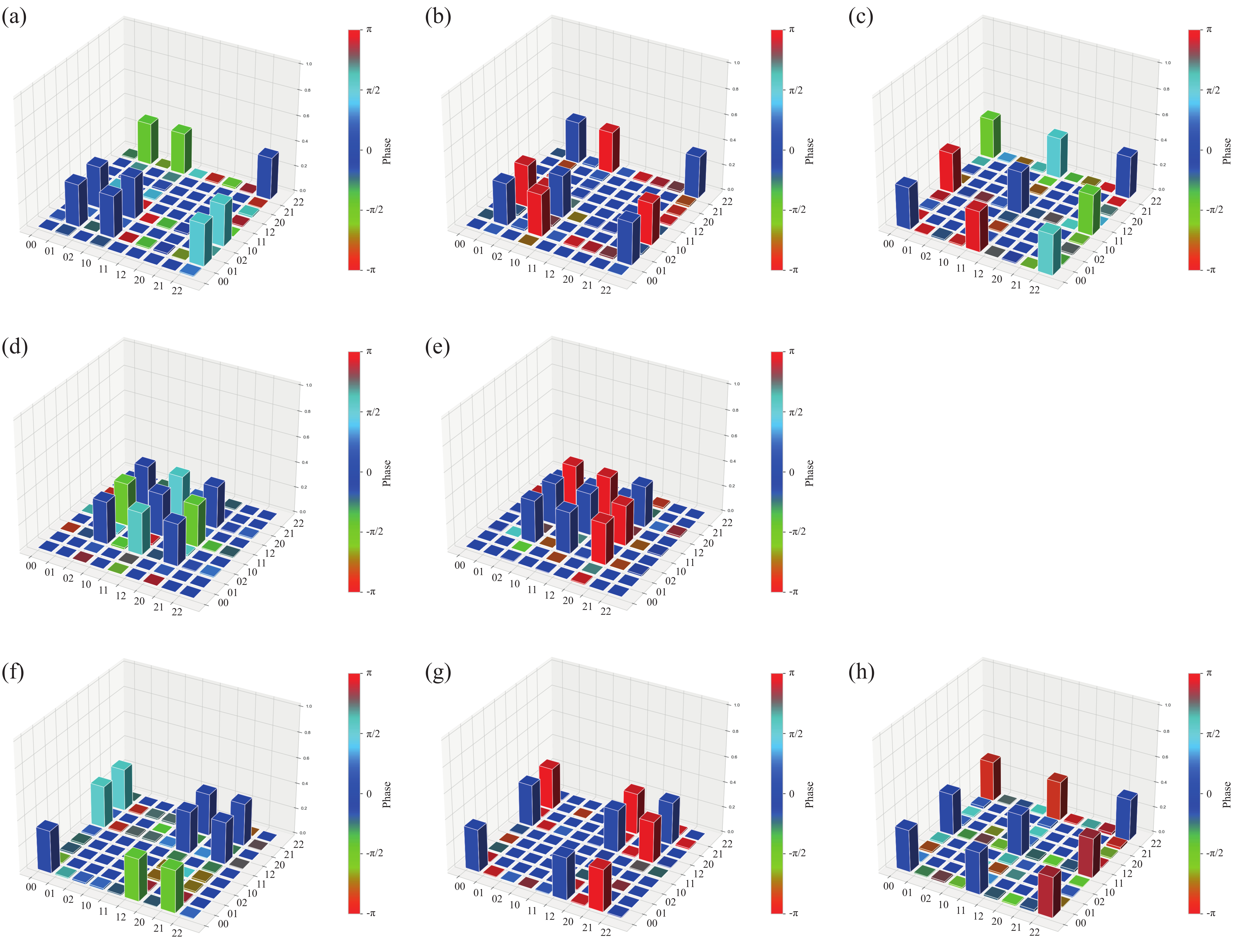}
    \caption{\textbf{Single qutrit Gell-Mann operations}. Shown are experimentally reconstructed Choi operators corresponding to the 8 Gell-Mann operations forming the basis for the qutrit Hilbert space. The fidelities are given in Tab.~\ref{tab:single_qudit_results}.}
    \label{fig:Supp_LocalGM}
\end{figure}

We performed quantum process tomography using Eq.~\eqref{eq:MLEoptQPT} and the basis in Eq.~\eqref{eq:QutritBasis} to reconstruct the single-qutrit Gell-Mann operators in Fig.~\ref{fig:Supp_LocalGM} and single-qutrit Clifford in Fig.~\ref{fig:Supp_LocalClifford}.

\begin{table}
\setlength\tabcolsep{1em}
\begin{tabular}{ c | l || c | l  }
\multicolumn{2}{c||}{Clifford + T} & \multicolumn{2}{|c}{Gell-Mann} \\
\hline\hline
$X_3$ & $\mathcal{F} = 0.994^{+0.003}_{-0.001}$ & $\lambda_1$ & $\mathcal{F} = 0.986^{+0.006}_{-0.001}$\\
$Z_3$ & $\mathcal{F} = 0.952^{+0.013}_{-0.017}$ & $\lambda_2$ & $\mathcal{F} = 0.992^{+0.003}_{-0.001}$\\
$S_3$ & $\mathcal{F} = 0.968^{+0.011}_{-0.012}$ & $\lambda_3$ & $\mathcal{F} = 0.967^{+0.017}_{-0.004}$\\
$H_3$ & $\mathcal{F} = 0.945^{+0.011}_{-0.015}$ & $\lambda_4$ & $\mathcal{F} = 0.989^{+0.004}_{-0.002}$\\
$T_3$ & $\mathcal{F} = 0.994^{+0.001}_{-0.016}$ & $\lambda_5$ & $\mathcal{F} = 0.970^{+0.012}_{-0.001}$\\
  &   & $\lambda_6$ & $\mathcal{F} = 0.991^{+0.004}_{-0.002}$\\
  &   & $\lambda_7$ & $\mathcal{F} = 0.985^{+0.008}_{-0.001}$\\
  &   & $\lambda_8$ & $\mathcal{F} = 0.960^{+0.016}_{-0.006}$
\end{tabular}
\caption{\label{tab:single_qudit_results}\textbf{Fidelities for single qutrit gates}. Fidelities are estimated from quantum process tomography without SPAM correction. The uncertainties correspond to $1\sigma$ statistical uncertainty from quantum projection noise, estimated using MonteCarlo resampling.}
\end{table}

\section{Single-qutrit results}

\end{document}